\documentclass[prd,showpacs,amsmath,amssymb,floatfix]{revtex4}
\usepackage{graphicx}

\begin{document}

\title{Energy Landscape of d-Dimensional Q-balls}

\author{Marcelo Gleiser and Joel Thorarinson}
\email{gleiser@dartmouth.edu; thorarinson@dartmouth.edu}

\affiliation{Department of Physics and Astronomy, Dartmouth College,
Hanover, NH 03755, USA}

\date{\today}

\begin{abstract}
We investigate the properties of $Q$-balls in $d$ spatial dimensions.  First, a
generalized virial relation for these objects is obtained. We then focus on
potentials  $V(\phi\phi^{\dagger})= \sum_{n=1}^{3} a_n(\phi\phi^{\dagger})^n$,
where $a_n$ is a constant and $n$ is an integer,
obtaining variational estimates for their energies
for arbitrary charge $Q$. These analytical estimates are contrasted with
numerical results and their accuracy evaluated. Based on the results, we offer
a simple criterion to classify ``large'' and ``small'' $d$-dimensional
$Q$-balls for this class of potentials. A minimum charge is then computed and
its dependence on spatial dimensionality is shown to scale as $Q_{\rm min} \sim
\exp(d)$. We also briefly investigate the existence of $Q$-clouds in $d$
dimensions.
\end{abstract}


\maketitle

\section{Introduction}

Many physical systems may be efficiently modeled in a reduced number of
spatial dimensions. In particular, certain static solutions of nonlinear partial
differential equations exhibiting solitonic behavior, or at least spatial 
localization, have been shown to exist from 
hydrodynamics and condensed matter physics \cite{h-cm} to relativistic
field theories \cite{rel}. At the opposite extreme, the possibility that the
four fundamental interactions of nature may be unified in theories with
extra spatial dimensions has triggered a search for static
nonperturbative higher-dimensional solutions of a variety of models derived
from string theory \cite{stringy-sol}. The extra dimensions may be compact and
much smaller than the usual three dimensions of space, as in Kaluza-Klein
models \cite{KK}, or they may be infinitely large, as in the recent 
Randall-Sundrum scenario, where gravity (and, possibly, other fields) can leak 
into one or more dimensions orthogonal to the 3-dimensional brane where matter
and gauge fields propagate and interact\cite{RS}. 

In either higher-dimensional
formulation, there is plenty of motivation to search for $d$-dimensional
nonperturbative configurations with $Q$ quanta. The reasons are two-fold:
first, as an upsurge
of recent work suggests, it is possible that these extra dimensions
may be much larger than the planckian scale \cite{Antoniadis,Csaki}. 
If the fundamental gravity scale is $M$, the associated
length scale of the extra dimensions is $R_{\rm KK}
\simeq (M_{\rm Pl}/M)^{2/(d-3)}M^{-1}$. 
[$d-3 \geq 1$ is the number of extra dimensions.] Thus, if $M\sim 1$ TeV,
$R_{\rm KK}\sim 10^{32/(d-3)} 10^{-17}$ cm. For $d\geq 5$, this scenario is 
still acceptable by current tests of Newton's gravitational law \cite{Rubakov}. 
In this case, it is possible that signatures of extra dimensions may be discovered
in collider experiments \cite{Antoniadis,Giudice, Quiros}. 
For example, energy may leak out into the bulk,
disappearing from three dimensional space (missing energy) or, 
more relevantly for the present
work, higher dimensional solitonic configurations may be produced at future
accelerators, as has been recently suggested by one of us \cite{d-oscillons}.
Thus, computing their properties, such as energy and spatial extension, and
how they scale with spatial dimensionality, is of great interest: if the 
fundamental interactions giving rise to the observed configuration are known,
their properties may serve as a probe to determine the dimensionality of
spacetime \cite{d-oscillons}.
Another reason for considering such objects is that
they may well exist in a pre-compactification spacetime. Being
nonperturbative, they may contribute in as yet unknown ways to the
compactification process or even to the stability of the compactified
internal space. Their existence may change the structure of the vacuum, a
possibility worth exploring in the near future.

In the present work we will focus on a particular class of nonperturbative
configurations called $Q$-balls. This choice reflects the fact that they
are -- as will be shown --
the simplest static, spatially-localized configurations that
can be found in any number of spatial dimensions. They have been shown
to appear in many versions of 3-dimensional field theories, 
supersymmetric or not (see references
in next paragraph) and will most certainly be present in 
their higher-dimensional
extensions, although we will not present a detailed study of specific
examples here. As in the early work concerning 3-dimensional $Q$-balls, 
our strategy is to start by
studying their properties in the simplest possible models that can support
them. More specific applications are the next logical step.
$Q$-balls are non-dissipative, spatially-localized scalar field configurations
that owe their stability to the conservation of a global charge $Q$
\cite{Coleman}. As such, they belong to the general class of nontopological
objects constructed to bypass the limitations imposed by Derrick's theorem
forbidding the existence of static localized configurations of real scalar fields
in more than one spatial dimension \cite{Derrick}. The distinguishing feature
of $Q$-balls is that they can be constructed from a single complex scalar field;
most nontopological solitons are built out of two or more interacting scalars
\cite{Lee}.

Since they were first proposed by Coleman, these objects have triggered many
different avenues of research: some searched for $Q$-balls in nonabelian models
\cite{nonabelian} or with several interacting fields \cite{gaugeQ},
others in supersymmetric models \cite{susyQ} and the effects of
their mutual interactions \cite{interacQ}, while others investigated
their stability against quantum
effects \cite{Graham} and
their cosmological role \cite{cosmoQ}. Understandably, this past work
was confined to the usual three spatial dimensions. Given the recent upsurge of
interest in higher-dimensional theories, we believe it important to investigate
how the properties of $Q$-balls depend on spatial dimensionality. Although
there is a trend to consider models with compactified extra dimensions,
in this work we will consider only $Q$-balls in flat $(d+1)$-dimensional
Minkowski spacetimes. One reason is that these objects will typically have
radii $R_Q$ which are much smaller than the scale of the compactified space,
$R_{KK}$, in particular in the context of models with large extra dimensions
\cite{Antoniadis}; that is, they will ``see'' a flat $d$-dimensional space.

Our emphasis throughout this work is to explore how spatial dimensionality
affects some of the key properties of $Q$-balls: their energy, minimal
charge, and size. For this purpose, we will employ a combination of
analytical and numerical tools, often contrasting one with another.

The paper is organized as follows. In section 2 we present the class of
models we will be working with, establishing notation and definitions. We also
obtain a $d$-dimensional virial relation for $Q$-balls that is shown to
encompass Derrick's theorem when $Q=0$. In section 3, we define the type of
potentials we will be using throughout this work and study
large $Q$-balls, using a variational ansatz
for their spatial profiles to estimate their energies and explore some of their
properties. We contrast our results with numerical computations. In section 4
we treat small $Q$-balls, again using a variational approach. We show that
the concept of ``small'' $Q$-ball makes more sense for lower dimensional models,
as dimensionality greatly affects the minimum charge for these objects. We obtain
an analytical estimate for the minimum charge $Q_{\rm min}$ as a function of
spatial dimension $d$ accurate for $d\leq 6$,
and find a simple relation, $Q_{\rm min}\sim \exp[d]$.
In section 5 we briefly explore the existence of $Q$-clouds in
$d$ dimensions. $Q$-clouds are unstable diffuse solutions of the equations
of motion discovered by Mark Alford a few years after Coleman first published
his paper \cite{Q-clouds}. We were unable to find $Q$-clouds for $d \geq 6$,
$d=5$ being marginal.
We close in section 6 presenting our conclusions and plans for future work.

\section{$Q$-ball Basics and $d$-Dimensional Virial Theorem}

\subsection{Basics}

Consider a scalar field $\phi$ with Lagrangian density

\begin{equation}
{\cal L} = \partial_{\mu}\phi^{\dagger}\partial^{\mu}\phi -
V(\phi\phi^{\dagger})~,
\label{lagrangian}
\end{equation}
where $\mu=0,1,...,d$ and the metric is $\eta_{\mu\nu} =
{\rm diag}(1,-1,-1...)$.
We will restrict our analysis to models with a conserved
$U(1)$ symmetry.

The potential $V(\phi\phi^{\dagger})$ is assumed to have a
global minimum at $\phi=0$. $Q$-balls are nonperturbative
excitations about this global vacuum state carrying a conserved charge $Q$,
the net particle number. Thus, if the energy of the $Q$-ball, $E_Q$, is
smaller than $Qm_{\phi}$ [$m^2_{\phi} = V''(0)$
is the mass of (perturbative)
particle-like vacuum excitations at $\phi=0$],
$Q$-balls are stable and energetically preferred. This
is an example of nonperturbative effects dominating the physics of the vacuum.
If a model supports such configurations, one cannot claim to understand the
physics of the vacuum before incorporating their contribution.
Configurations with lowest
energy are obtained by writing \cite{Coleman, Lee}

\begin{equation}
\phi({\bf x},t) = \frac{1}{\sqrt{2}}
\Phi({\bf x})e^{i\omega t}~.
\label{scalarfield}
\end{equation}

From the associated conserved current
$j^{\mu} = -i(\phi^{\dagger}\partial^{\mu}\phi -
\phi\partial^{\mu}\phi^{\dagger})$, one obtains the charge conservation law,

\begin{equation}
Q = \omega \int d^dx \Phi^2~.
\label{conservedQ}
\end{equation}

In practice, $Q$-balls are very similar to critical bubbles in the theory
of homogeneous nucleation or, to use Coleman's expression, false vacuum decay
\cite{falsevacuum}. The
latter is characterized by an asymmetric double-well potential, with the
field initially trapped in the false vacuum state. The critical bubble or
bounce is the spherically-symmetric
static solution to the equation of motion that interpolates
(or almost) between the two potential minima: at $r=0$, $\phi(r=0)\equiv \phi_c$
is close to the global
minimum $\phi_+$
and, as $r\rightarrow\infty$, it approaches the false vacuum at $\phi=0$. The field
configurations for $Q$-balls do the same. But how could this be if the
potential $V$ has a global and not a local
minimum at $\phi=0$? The answer, a clever trick,
comes from the term in the equation of motion
proportional to $\ddot\phi \propto -\omega^2\Phi$.
This term contributes an effective negative mass to the potential:
for certain
choices of $\omega$, it makes the minimum at some
$\phi_+$ the global minimum and
the minimum at $\phi=0$ the false vacuum. One can see this by varying the
action obtained from the Lagrangian in eq. \ref{lagrangian} using the
form of the field of eq. \ref{scalarfield} to obtain,

\begin{equation}
\nabla_d^2\Phi = -\omega^2\Phi +\frac{\partial V}{\partial \Phi}
\equiv  U'(\Phi)~.
\label{eqofmotion1}
\end{equation}

The particle ``moves'' in the effective potential $-U(\Phi)$. As $\omega
\rightarrow m_{\phi}$ the barrier separating the two minima disappears
and the vacuum at $\phi=0$ becomes classically (spinodally) unstable.

\subsection{Virial Theorem}

From the Lagrangian density of eq. \ref{lagrangian} and using eq.
\ref{scalarfield}, we obtain the energy
for a field configuration in $d$ spatial dimensions as

\begin{equation}
E[\Phi] = \int d^dx \left [\frac{1}{2}(\nabla_d\Phi)^2 +\frac{1}{2}
\omega^2\Phi^2 + V(\Phi^2)\right ]~.
\end{equation}

We can use eq. \ref{conservedQ} to express the energy in terms of the
conserved charge $Q$,

\begin{equation}
E_Q[\Phi] = \frac{1}{2}\frac{Q^2}{\langle \Phi^2\rangle } + \langle \frac{1}{2}
(\nabla_d\Phi)^2\rangle + \langle V\rangle~,
\label{EQ}
\end{equation}
where $\langle \dots \rangle \equiv \int \dots d^dx$.
Now scale the spatial variable
$x \rightarrow \alpha x$ and impose that the energy is invariant under
scale transformations, $\partial E/\partial \alpha|_{\alpha=1} = 0$, to obtain
the virial relation,

\begin{equation}
d\langle V\rangle = (2-d)\langle \frac{1}{2}(\nabla_d\Phi)^2\rangle +
\frac{d}{2}\frac{Q^2}{\langle \Phi^2\rangle }~.
\label{virial}
\end{equation}

This is a generalization of Derrick's theorem for $Q$-balls in an arbitrary
number of dimensions. [See ref. \cite{Kusenko97} for a derivation in $d=3$.]
Derrick's theorem is easily recovered for $Q=0$. We can
also see how $Q$-balls can evade it: for $d\geq 2$, the contribution from the
charge term, being positive definite, may compensate the negativity of the
surface term. In fact, an absolute lower bound for $Q$-balls to be preferred
energy state is

\begin{equation}
Q^2 \geq \frac{2(d-2)}{d}\langle \Phi^2\rangle
\langle \frac{1}{2}(\nabla_d\Phi)^2\rangle~.
\label{Qcond}
\end{equation}
Note that for $d=2$ there is no condition on the minimum charge.

Using the virial relation, we can eliminate the potential or charge
term from the expression of the energy, eq. \ref{EQ}, to obtain
either of the two following expressions for energy density,

\begin{equation}
\frac{E}{Q} = \omega +\frac{2}{d}\frac{\cal S}{Q} = \omega \left\{1
+\left(\frac{1}{d-2+d \frac{\cal V}{\cal S}} \right) \right\} \leq
m_{\phi}~, \label{Econd1}
\end{equation}
where we introduced the shorthand notation,
${\cal S} \equiv \langle\frac{1}{2}(\nabla_d\Phi)^2\rangle$ and
${\cal V} \equiv \langle V(\Phi)\rangle$.
The
inequality reflects the condition that for $Q$-balls to be the
favored energy state, $E/Q \leq m_{\phi}$. One can see that this
condition is consistent with the fact that $\omega <m_{\phi}$ for a
solution to exist: otherwise the valley at the origin becomes a hill
\cite{Coleman, Q-clouds}. Also, $Q$-balls are favored for large
charges and small surface energy.

\section{Large Q-balls: Variational Approach}

Large $Q$-balls are characterized by a large charge and radius.
We will
explore their general properties (including what is meant by ``large'')
in an arbitrary number of spatial dimensions
using a variational approach in the spirit of Friedberg, Lee, and Sirlin's
treatment of 3-dimensional nontopological solitons \cite{Friedberg}.
In his original $Q$-ball paper, Coleman presented a simple estimate for the
energy of a large 3d $Q$-ball by neglecting the surface term and taking
the field to be constant inside the ball's volume $v$ \cite{Coleman}.
From eq. \ref{EQ},
it's easy to write the energy as

\begin{equation}
E = \frac{1}{2}\frac{Q^2}{\Phi_c^2v} + V(\Phi_c)v~.
\label{colemanestimate}
\end{equation}

Extremizing with respect to the volume $v$, one obtains

\begin{equation}
E = Q\sqrt{2V(\Phi_c)/\Phi_c^2}~.
\label{colemanenergy}
\end{equation}
We see that, within this approximation, for $Q$-balls to exist,
$\sqrt{2V(\Phi_c)/\Phi_c^2} < m_{\phi}$. At this point
we will choose
a specific polynomial form for the potential $V(\phi)$, which we will adopt
throughout this work. Our results will hold for this choice.
Other potentials, such as those in ref. \cite{Kusenko97}, which contains
terms of the form  $\sim (\phi\phi^{\dagger})^{3/2} $ require a 
separate study. Accordingly, we write

\begin{equation}
V(\phi\phi^{\dagger}) = m^2\phi\phi^{\dagger} - b(\phi\phi^{\dagger})^2
+\frac{4c}{3}(\phi\phi^{\dagger})^3~,
\label{potential1}
\end{equation}
where the constants $m^2$, $b$, and $c$ are real and positive. Writing the
field as in eq. \ref{scalarfield} and introducing the dimensionless
field $X^2 \equiv \sqrt{c/m^2}\Phi^2$, angular frequency $\omega' = \omega/m$,
and spacetime variables $x_{\mu}' = x_{\mu}m$, the energy
of a configuration can be written as

\begin{equation}
E[X] = \frac{m^{3-d}}{\sqrt{c}}\int d^dx'\left [\frac{1}{2}\omega'^2X^2
+\frac{1}{2}(\nabla_d^{'2} X)^2 +\frac{X^2}{2} - \frac{b'}{4}X^4 + \frac{X^6}{6}
\right ]~,
\label{energy}
\end{equation}
where $b' \equiv b/(mc^{1/2})$. Notice that in order for the symmetric minimum
of $V(X=0)$ to remain the global minimum $b' \leq 4\sqrt{3}/3\simeq 2.309$.

\begin{figure}
\includegraphics[width=3in,height=3in]{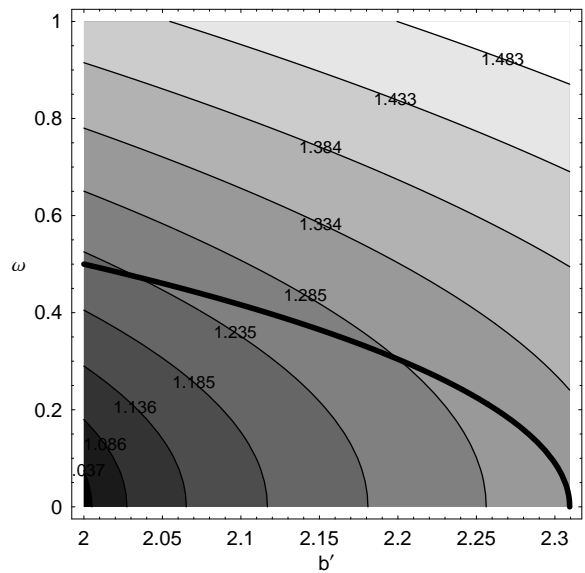}
\caption {Surface plot showing values of $X_+$, the local minimum of
$V(X^2)$, as a function of $\omega$ and $b'$. The bands denote
degenerate values in decrements of $0.05$ starting at the top from
$1.483$. The solid line represents eq. \ref{omega}.} \label{plotV}
\end{figure}

In fig. \ref{plotV} we plot $X_+$, the local minimum of $V(X^2)$, as
a function of $\omega$ and $b'$. The surface plot displays the bands
with degenerate values of $V(X_+)$ for pairs of $(\omega,b')$. The
bottom line
represents the inequality
\begin{equation}
\omega_c \geq \sqrt{1-\frac{3 b'^2}{16}}. \label{omega}
\end{equation}
This inequality comes from imposing that $U(X=0)\geq
U(X_+)$ so that a solution for eq. \ref{eqofmotion1} interpolating
between the two minima (the ``bounce'') satisfying the correct
boundary conditions exists \cite{falsevacuum}.
For large $Q$-balls, $X_+$ is the approximate value of
the field at the core of the $Q$-ball, $X_c$. Once surface terms are
included, the shape degeneracy of eq. \ref{colemanenergy} is broken
and spherical symmetry is favored. Hence, from now on we will only
consider spherically-symmetric $Q$-balls. We will also drop the
primes (except for $b'$) and work with dimensionless variables. Note that with this
choice of potential, $V''(0)= m^2_{\phi}=1$. This is the value that
should be used on the inequality of eq. \ref{Econd1}.

\begin{figure}
\resizebox{80mm}{!}{\includegraphics{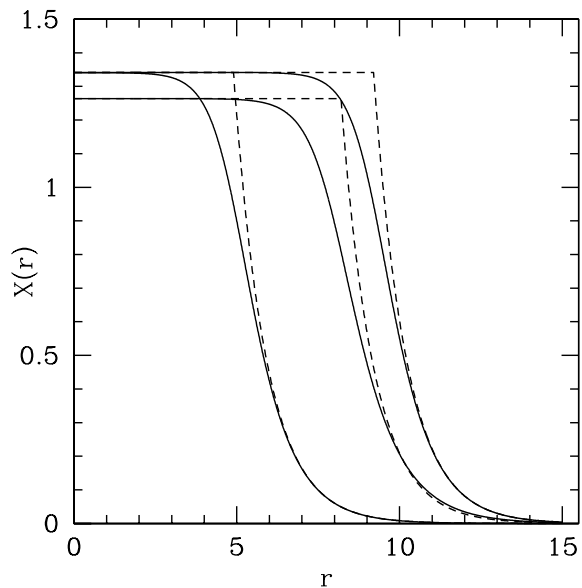}} \caption {A sample
of $Q$-ball profiles obtained numerically for various choices of
$(\omega,~b',~d)$. From left to right $(0.8,2.0,5), (0.6,2.0,3),
(0.8,2.0,8)$. The dotted lines are the equivalent profiles from the
ansatz of eq. \ref{ansatz1}.} \label{numX}
\end{figure}

In fig. \ref{numX}
we show a sample of numerical solutions for $X(r)$ for choices of $\omega,~b'$,
and $d$,
which are well-approximated by the large $Q$-ball ansatz to be presented below.
In fig. \ref{numerQ}, we plot the numerically-evaluated $Q$-ball
energy as a function of charge for different spatial
dimensions. The left-hand side plot is for $b'=2$ and the right hand side for $b'=2.3$,
covering the allowed range of values of $b'$. The
energy was obtained by first solving the equation for $\Phi(r)$ numerically
(eq. \ref{eqofmotion1}, see fig. \ref{numX})
and then using the result in eq. \ref{energy}. There are several
points to note: i) the existence of a minimum charge, $Q_{\rm min}$,
below which $E/Q>1$;
ii) the existence of
$Q$-balls in higher dimensions and with considerably larger values of the
charge $Q$; iii)
the existence of a cusp in the region $E/Q>1$ for $d=3,~4,~5$
with a second branch, that is, the existence of
different configurations with the same value of the charge.
These are the $Q$-clouds described in ref. \cite{Q-clouds}, which we were able to
find for
$d\leq 5$. We will come back to these three points below.

\begin{figure}
    \centering
    \begin{minipage}[b]{0.5\textwidth}
\resizebox{80mm}{!}{\includegraphics{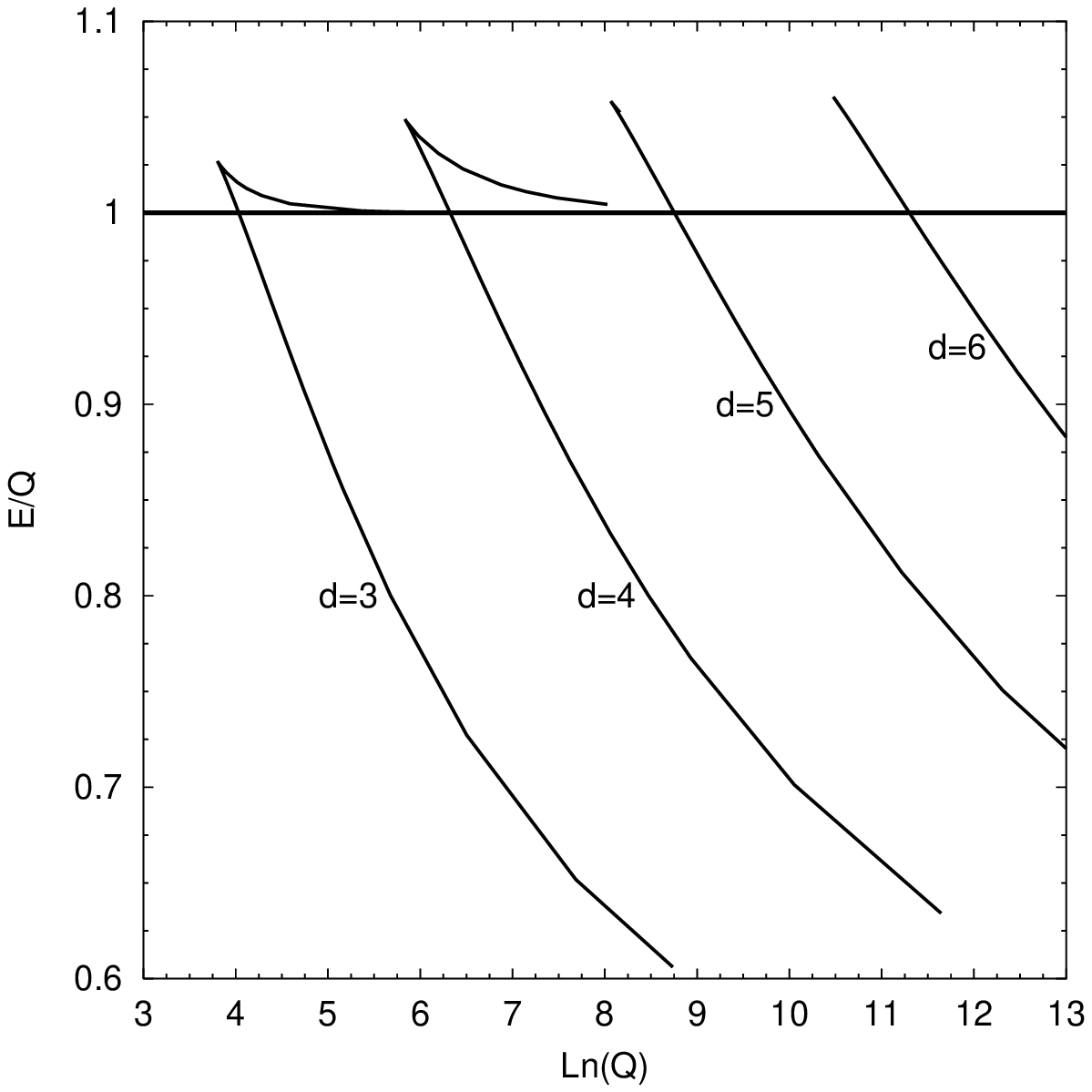}}
    \end{minipage}%
\begin{minipage}[b]{0.5\textwidth}
\resizebox{80mm}{!}{\includegraphics{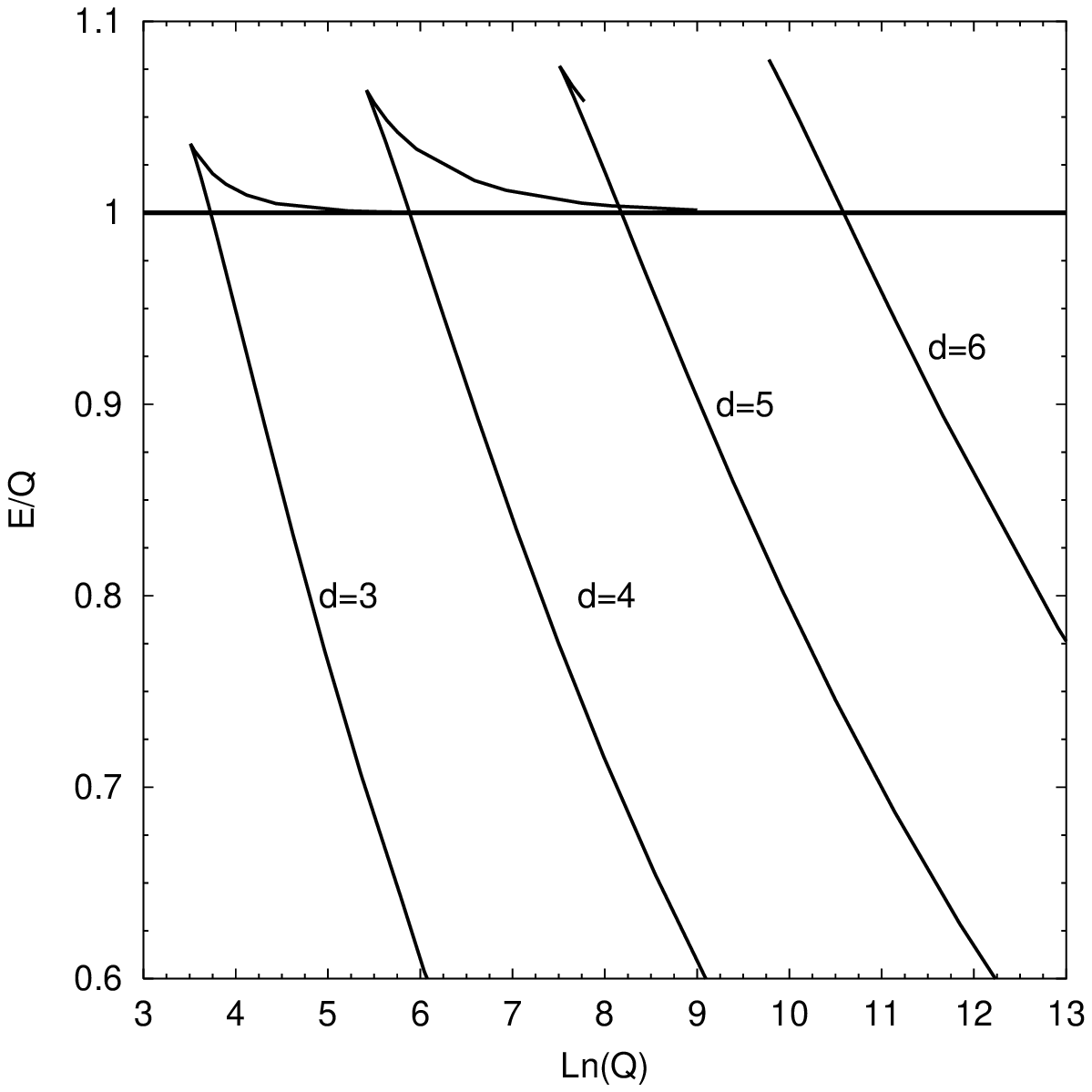}}
    \end{minipage}%
\caption {{$E/Q$ as a function of charge $Q$ for various dimensions.
$b'=2$ on the left and $b'=2.3$ on the right.}} \label{numerQ}
\end{figure}

Choosing a variational profile for the field has a degree of
arbitrariness. Our guiding principle was a compromise between
accuracy and calculability. Unfortunately, there is a sort of
uncertainty principle operating here, in that higher accuracy
implies in lower calculability and vice-versa. The works of refs.
\cite{Woods} used a Woods-Saxon ansatz for $d=3$. We verified that this ansatz is
also quite accurate for $d\geq3$ (matching closely the numerical
solutions) but not as useful for the kind of analytical calculations
we are interested in. We thus opted for the following simple ansatz
for the field profile:

\begin{eqnarray}
 X(r) = \left\{ \begin{array}{ll}
          X_c & \mbox{ if $r < R$} \\
      X_c \exp[-\alpha(r-R)] & \mbox{ if $r\geq R$},
               \end{array}
       \right.
\label{ansatz1}
\end{eqnarray}
where $\alpha$ is a variational parameter to be determined.

Applying this ansatz to eq. \ref{energy} one obtains, after some algebra,

\begin{eqnarray}
E[X] & \leq & \frac{c_d}{d}\left [\frac{1}{2}\omega^2X_c^2+ V(X_c^2)\right ]R^d\nonumber \\
     &  & + c_d(d-1)!\sum_{k=0}^{d-1}\frac{1}{k!}\left [\frac{A}{(2\alpha)^{(d-k)}}
     -\frac{B}{(4\alpha)^{(d-k)}}+\frac{C}{(6\alpha)^{(d-k)}}\right ]R^k~,
\label{energy2}
\end{eqnarray}
where,
\begin{equation}
A\equiv \frac{X_c^2}{2}(\omega^2+\alpha^2+1);~~B\equiv \frac{b'}{4}X_c^4;~~C\equiv \frac{X_c^6}{6}~
\label{defsABC}
\end{equation}
and we have used that $\int d^dx \rightarrow c_d\int r^{(d-1)}dr$. $c_d$ is related to the
volume of the $d$-dimensional sphere, $v_d = c_dr^d$, with
$c_d \equiv 2\pi^{(d/2)}/\Gamma(d/2)$. Using the ansatz, one can also obtain the conserved
charge $Q$,
\begin{equation}
Q = \omega c_d X_c^2\left [\frac{R^d}{d} + (d-1)!\sum_{k=0}^{(d-1)}\frac{r^k}{k!(2\alpha)^{(d-k)}}
\right ]~.
\label{Qlarge}
\end{equation}

A quick glance at the expressions for the energy and conserved charge and we can see that to
proceed we must make simplifications. We will include only the dominant term from the
surface contribution, proportional to $ R^{(d-1)}$. As we will see,
this approximation is quite accurate for large
$Q$-balls. Using the expression for $Q$ in eq. \ref{energy2} we obtain,

\begin{equation}
E_Q\leq \beta_Q^2R^{-d} + \frac{c_d}{d}V(X_c^2)R^d + Zc_d R^{(d-1)}~,
\label{energy3}
\end{equation}
where we introduced the notation
\begin{equation}
\beta_Q^2\equiv \frac{dQ^2}{2c_d X_c^2}~~; Z\equiv \frac{1}{2\alpha}
\left [ (1+\alpha^2)\frac{X_c^2}{2}-\frac{B}{2} + \frac{C}{3}\right ]~.
\label{betaZ}
\end{equation}

Before we extremise eq. \ref{energy3}
with respect to $R$, we extend Coleman's result to $d$ dimensions, that is,
we extremise the energy
neglecting the surface term to obtain the zeroth-order approximation
for the $Q$-ball radius,
\begin{equation}
R_0^{2d} = \frac{d\beta_Q^2}{c_d V(X_c^2)}~.
\label{zerothR}
\end{equation}

Substituting
$R_0$ into the expression for the energy (without surface term) we reproduce,
as we should, Coleman's result,
$E_{R_0}[X_c^2] = \sqrt{2V(X_c^2)/X_c^2}Q$: within this
approximation, $Q$-balls exist if $\sqrt{2V(X_c^2)/X_c^2} \leq 1$. Note that
this approximation provides no information on the role of spatial dimensionality
on the stability of $Q$-balls.

Imposing that $\partial E/\partial R|R_c = 0$ into eq. \ref{energy3} leads to the polynomial
equation
\begin{equation}
c_dZ(d-1)R_c^{(2d-1)} + c_d V(X_c^2)R_c^{(2d)} - d\beta_Q^2 = 0~.
\label{rcrit}
\end{equation}

In order to find an approximate solution, valid for large $Q$-balls, we write the critical
radius as $R_c \simeq  R_0 +\delta R$, where $\delta R \ll R_0$. Substitution into the
above equation
leads to an expression for $\delta R$ and thus for $R_c$,
\begin{equation}
\delta R = -\frac{(d-1)Z}{2dV(X_c^2)}~.
\label{deltar}
\end{equation}

We now substitute this approximate expression for $R_c$ into the expression for the energy,
eq. \ref{energy3}, to obtain,

\begin{equation}
E_Q[X_c]|_{R_c} = \sqrt{2V(X_c^2)/X_c^2}Q\left [1 + \xi Q^{-1/d}\right ]~,
\label{energy4}
\end{equation}
where we introduced
\begin{equation}
\xi \equiv \frac{dZ}{2}\left (\frac{c_d}{d}\right )^{1/d}\left [\frac{X_c^2}{\left [V(X_c^2)
\right ]^{(2d-1)}} \right ]^{(1/2d)}~.
\label{xi}
\end{equation}

This expression for the energy depends still on the variational parameters $X_c$ and, through
the expression for $Z$ (cf. eq. \ref{betaZ}), on $\alpha$. Ideally, one would extremise it for
these two variables to obtain a final expression for the energy given only in terms of the
spatial dimensionality $d$ and the asymmetry parameter $b'$. Although this task cannot be
done exactly, it can be done approximately. The final test for the efficacy of the approximations
will be to compare the analytical expression of the energy
to its numerical evaluation shown in fig. \ref{numerQ}. An estimate for $X_c$ can
be obtained by extremizing the energy neglecting the surface term, that is, by extremizing
Coleman's expression $E_{R_0}[X_c^2] = \sqrt{2V(X_c^2)/X_c^2}Q$. One obtains,

\begin{equation}
{\bar X}_c^2 = \frac{3}{4}b'~.
\label{Xcrit}
\end{equation}

Although, to quote Coleman \cite{Coleman}, neglecting the surface term amounts to a ``brutal set of
approximations,'' this analytical result for the core value of the field proves to be
quite reasonable. This
is due to the fact that $X_c$ is only mildly dependent on $d$ and $b'$.

In order to extremise $\alpha$ we revert to the expression for the
energy in eq. \ref{energy3}, and note that $\partial E/\partial\alpha = \partial Z/\partial\alpha$,
where $Z$ is defined in eq. \ref{betaZ}. One can easily obtain that

\begin{equation}
\alpha_c^2 = 1 - \frac{b'}{4}X_c^2+\frac{X_c^4}{9} = 1 - \frac{b'^2}{8}~,
\label{alphac}
\end{equation}
where in the last step we used eq. \ref{Xcrit}. Using these results in eq. \ref{energy4},
we obtain the promised
extremised expression for the $Q$-ball energy in terms of $d$ and $b'$,

\begin{equation}
\frac{E_Q[d,b']}{Q} \simeq \sqrt{1-\frac{3}{16}b'^2} \left [ 1 + \xi(d,b')Q^{-1/d} \right ]~,
\label{energy5}
\end{equation}
where,

\begin{equation}
\xi(d,b') = \frac{3d}{16}\left (\frac{c_d}{d}\right )^{1/d}b'\sqrt{1-\frac{b'^2}{8}}
\left [\frac{3b'}{4 \left [\frac{3b'}{8}\left (1-\frac{3b'^2}{16}\right )\right ]^{2d-1}}
\right ]^{1/2d}~.
\label{xi2}
\end{equation}
Glancing at eq. \ref{energy5}, we can see that since the
contribution from the surface term is positive, the lower bound on
the $Q$-ball energy is $E/Q > \sqrt{1-\frac{3}{16}b'^2}$, which
interestingly, is the same as saying $E/Q > \omega_c$ where $\omega_c$,
defined in eq. \ref{omega}, is
the minimum value that gives a solution to eq.
\ref{eqofmotion1}. Note that as
$b'^2\rightarrow 16/3$, which is the critical value for $V(\Phi=0)$
to be the global minimum, the term proportional to $Q^{-1/d}$
dominates the expression for $E/Q$, which is finite for all $b'$.

\begin{figure}
\centering \resizebox{80mm}{!}{\includegraphics{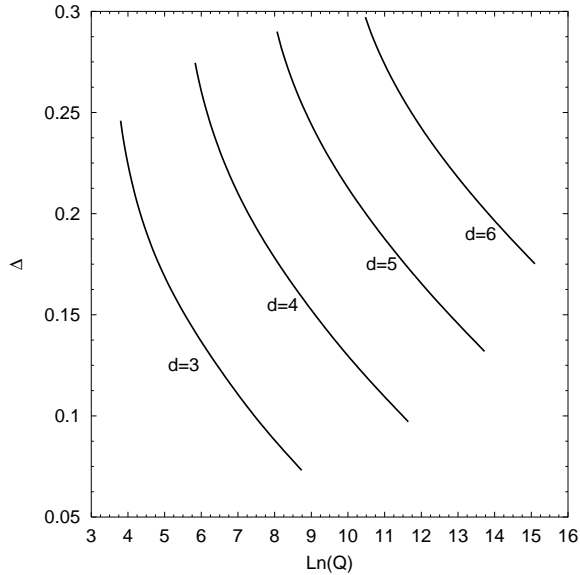}} \caption
{Contrasting the numerical and analytical expressions for the energy
as a function of charge $Q$ for various dimensions and $b'=2.0$.
$\Delta\equiv \left(|E/Q|_{\rm num} - E/Q|_{\rm
anal}|\right)/E/Q|_{\rm num}$. Note that, although not shown, 
the lines continue to
decrease monotonically as $Q$ increases, and the approximation improves.}
\label{DeltaQplot}
\end{figure}

In fig. \ref{DeltaQplot} we contrast the numerical results (see fig. \ref{numerQ}) with those
obtained from the analytical approximation
of eq. \ref{energy5} for different
values of $b'$ and $d$ as a function of $Q$. For clarity, we introduced the
percentual difference
$\Delta\equiv \left(E/Q|_{\rm num} - E/Q|_{\rm anal}\right)/E/Q|_{\rm num}$.
We plot $\Delta$ vs. $Q$,
showing that there is a wide range of parameter space where the approximation is better than
$20\%$. The agreement gets worse as $b'$ and $d$ increase,
although it always improves for larger values of $Q$ as one
would expect from an approximation describing large $Q$ balls. We will soon
offer a criterion to define what is a large or small $Q$-ball.

\section{Small Q-Balls: Variational Approach}

\begin{figure}
    \centering
    \begin{minipage}[b]{0.5\textwidth}
\resizebox{80mm}{!}{\includegraphics{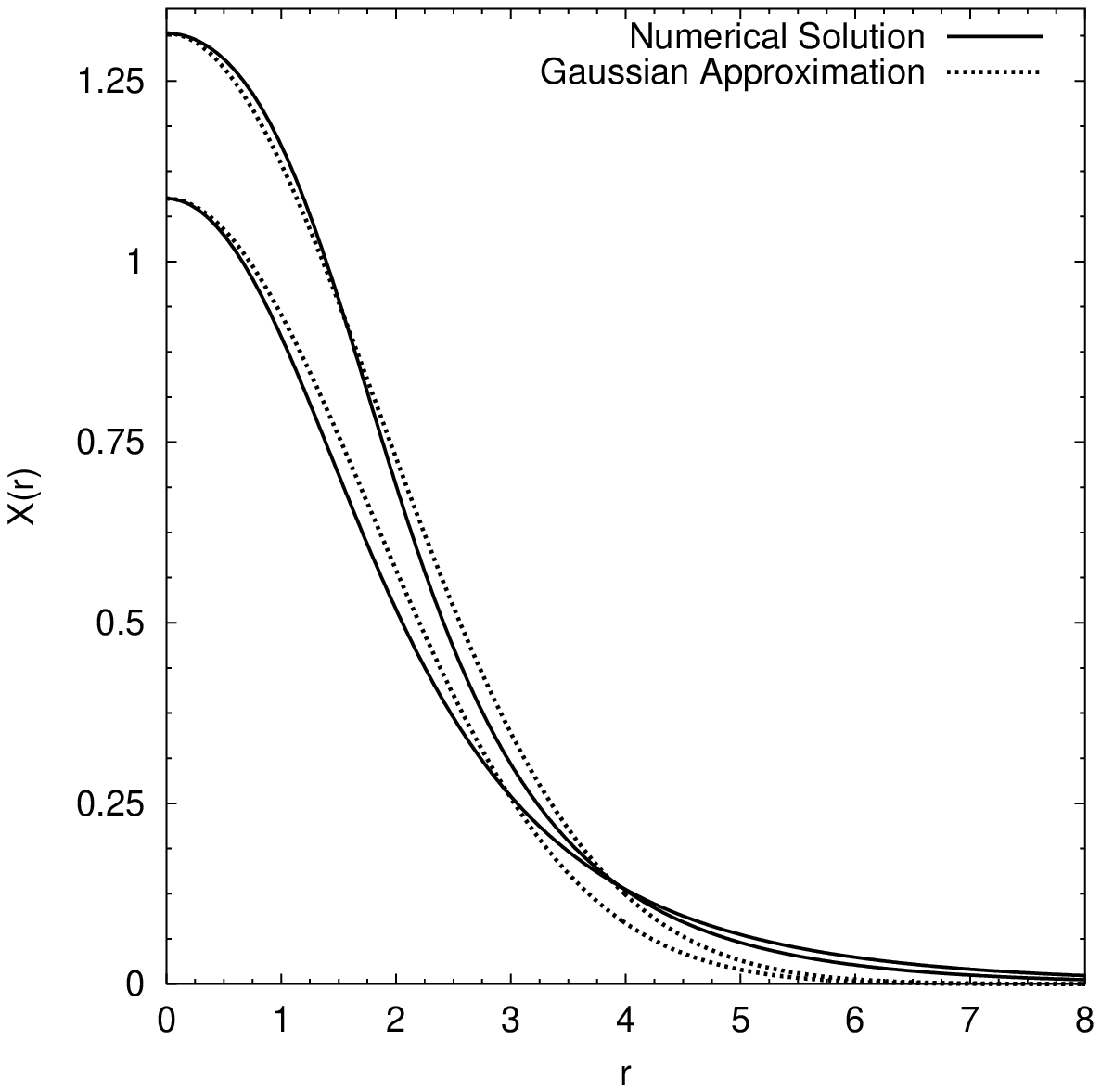}}
    \end{minipage}%
\begin{minipage}[b]{0.5\textwidth}
\resizebox{80mm}{!}{\includegraphics{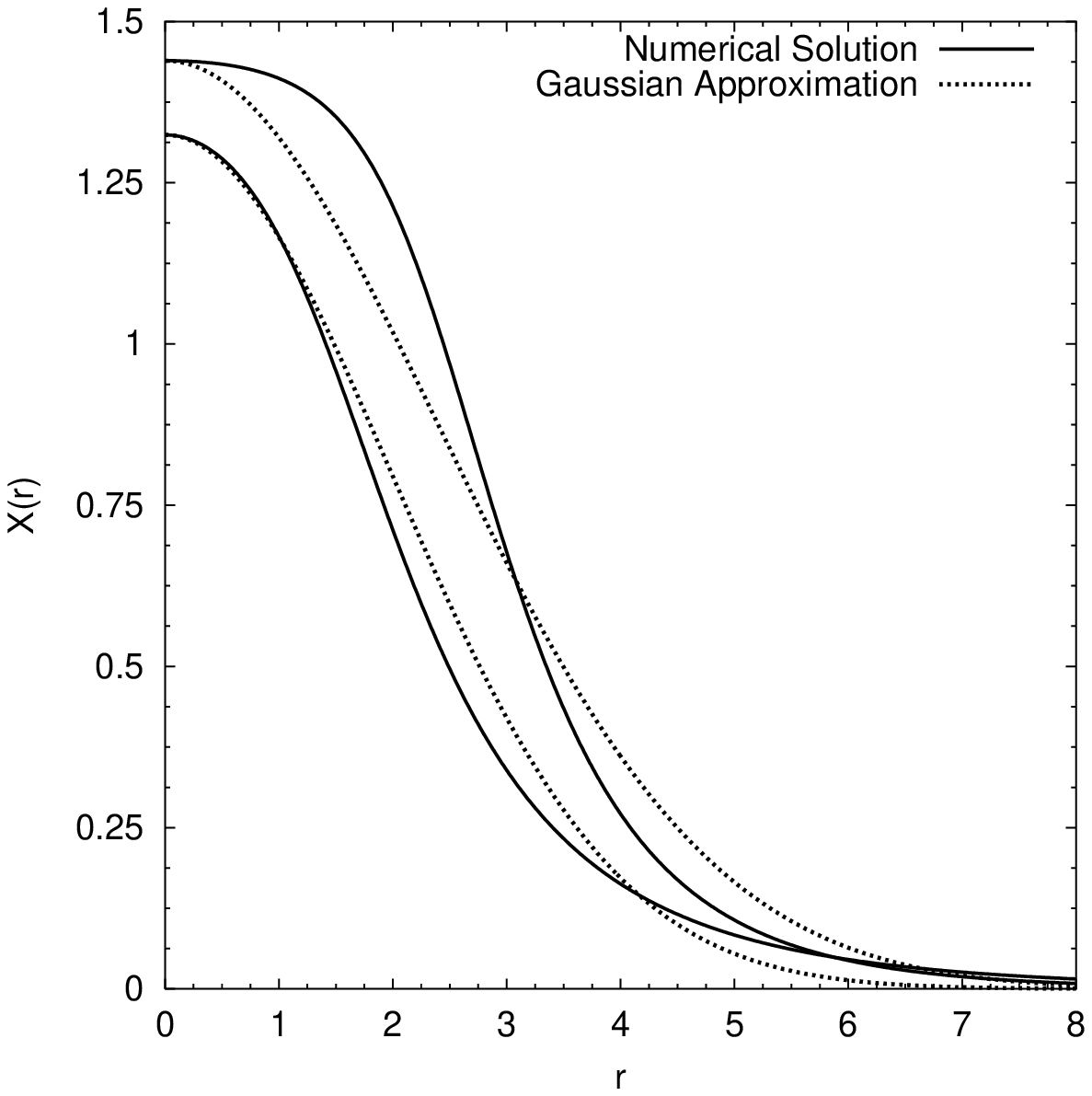}}
    \end{minipage}%
\caption {Comparison between numerical solution and gaussian profile
for a few sample ``small'' $Q$-balls. $b'=2.3$ throughout. On the
left $d=3$ and on the right $d=4$.The solid lines are the numerical
solutions to eq. \ref{eqofmotion1}, while the dashed lines are the
gaussian approximations of eq. \ref{gaussian}. From left to right,
$Q= 33,45,225,519$.}\label{gaussianfield}
\end{figure}

$Q$-balls can also be ``small,'' that is, not well-described by the
widely adopted thin-wall approximation, consistent with the results
above. Small $Q$-balls will have radii $R\gtrsim m^{-1}_{\phi}$. A
sample of such configurations are shown in fig. \ref{gaussianfield}.

In order to obtain analytical insight into these objects, we will
adopt a gaussian ansatz for their profile,

\begin{equation}
X(r) = X_c\exp[-r^2/R^2]~.
\label{gaussian}
\end{equation}
This approach, in contrast to that of ref. \cite{Kusenko97} which used an approximate
3d bounce solution, is particularly simple to adopt in an arbitrary number of spatial
dimensions. Using the gaussian profile above in the
expression for the energy eq. \ref{energy} and using dimensionless variables, we
obtain,

\begin{equation}
E_Q[X_c,R] = \left (\frac{\pi}{2}\right )^{d/2}\left [\beta_Q^2R^{-d}+\alpha R^{(d-2)} +\gamma
R^d \right ]~,
\label{energyg1}
\end{equation}
where we introduced,
\begin{equation}
\beta_Q^2 = \left (\frac{2}{\pi}\right )^d\frac{Q^2}{X_c^2};~~\alpha = dX_c^2;~~\gamma =
\frac{X_c^2}{2} - \frac{b'}{2^{d/2}}\frac{X_c^4}{4}+\frac{1}{3^{d/2}}\frac{X_c^6}{6}~.
\label{defsg1}
\end{equation}
The conserved charge is,

\begin{equation}
Q = \left (\frac{\pi}{2}\right )^{d/2}\omega X_c^2R^d~.
\label{chargeg1}
\end{equation}

Note that with the gaussian ansatz, it is easy to obtain the energy for arbitrary
polynomial potentials \cite{d-oscillons}. Writing $V(\Phi) = \sum_{n=1}^h \frac{g_n}{n!}\Phi^n$,
and $\Phi = A\exp[-r^2/R^2]+B$ ($B$ is the asymptotic vacuum), we get, $\int r^{(d-1)}V(\Phi)dr =
\frac{\Gamma(d/2)}{2}R^d\sum_{n=1}^h\frac{A^n}{n^{d/2}n!}\frac{d^nV}{d\Phi^n}|_B$.

By extremizing the expression for the energy, and adopting the same approximation
as in the previous section ($R_c = R_0+\delta R$), one computes the critical radius,

\begin{equation}
R_c = R_0\left [1- \frac{(d-2)}{dR_0^2}\frac{\alpha}{\gamma}\right ]~,
\label{Rcritg}
\end{equation}
where $R_0$ is the zeroth-order result, obtained by neglecting the surface term (proportional
to $\alpha$ in eq. \ref{energyg1}, exact for $d=2$),

\begin{equation}
R_0^{2d} = \frac{\beta_Q^2}{\gamma}~.
\label{zerothrg}
\end{equation}
Including the surface term decreases the radius as it should. The next step is to substitute
this critical radius back into the expression for the energy, eq. \ref{energyg1}.
We found it useful to write $R_c = R_0(1-\varepsilon)$ to keep track
of the radial corrections induced by the surface term (proportional to $\alpha$).
The final result is

\begin{equation}
E_Q[X_c] = \frac{\gamma^{1/2}}{X_c}Q\left [2 +\xi Q^{-2/d} - \frac{(d-2)^2}{d}\xi^2 Q^{-4/d}
\right ]~,
\label{energyg2}
\end{equation}
where,

\begin{equation}
\xi = \frac{\pi}{2}\frac{\alpha}{\gamma^{(d-1)/d}}X_c^{2/d}~.
\label{xig}
\end{equation}
This way of expressing the result makes it clear what the zeroth-order value for the energy
is ($\xi=0$)
and the role of surface corrections in arbitrary dimensions.
It is also quite useful to test the validity of the gaussian
approximation in the computation of the critical charge $Q_{\rm min}$, below which $E/Q >1$,
(see fig. \ref{numerQ}).
We use the value of $X_c$ in terms of $b'$ given by eq. \ref{Xcrit}, since it proves to be
quite accurate (cf. figs. \ref{plotV} and \ref{gaussianfield}).
Thus, $\alpha$, $\gamma$, and $\beta_Q^2$ can be expressed
in terms of $d$, $b'$, and $Q$ as in the previous section.

\begin{figure}
    \centering
    \begin{minipage}[b]{0.5\textwidth}
\resizebox{80mm}{!}{\includegraphics{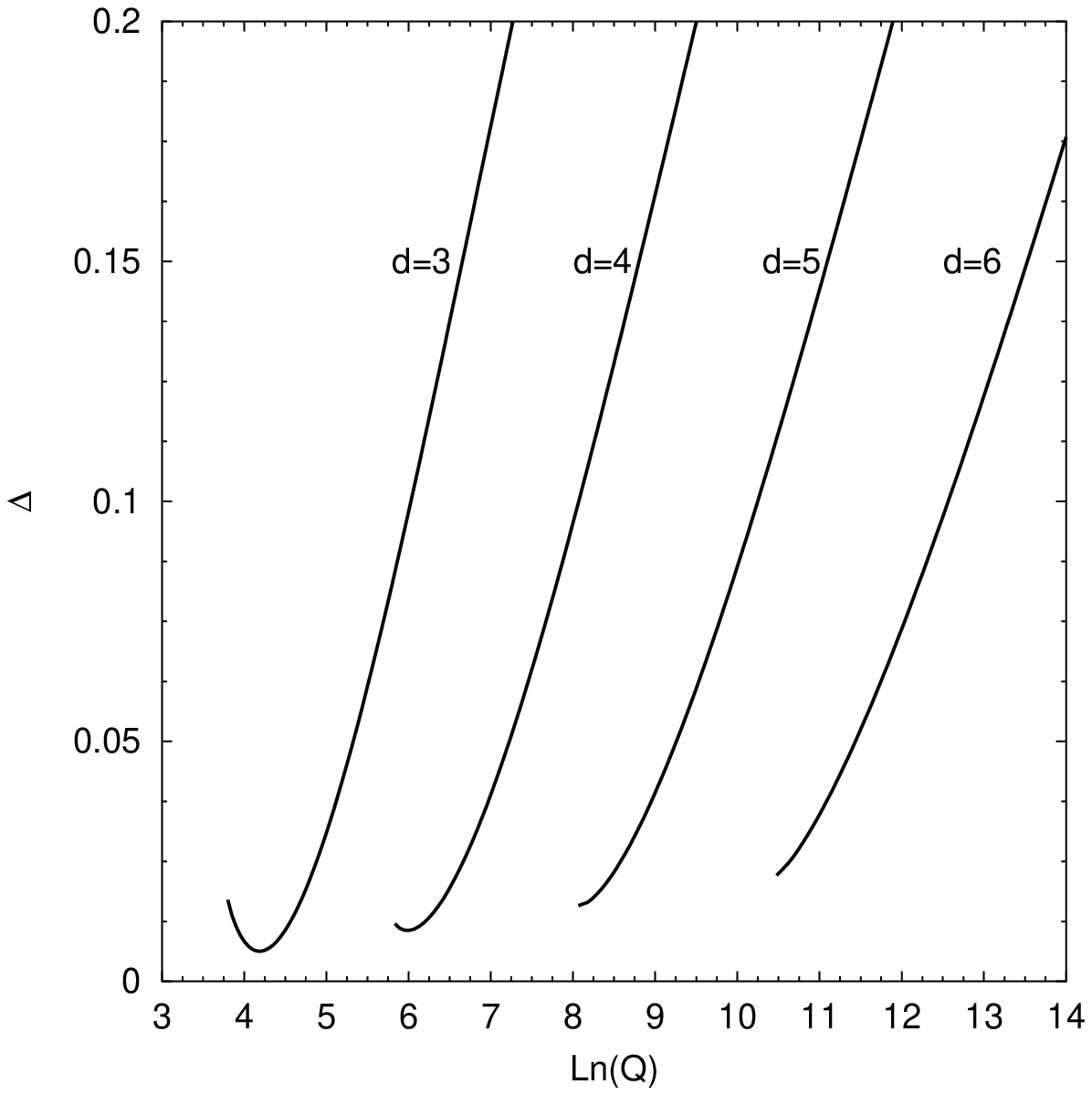}}
    \end{minipage}%
\begin{minipage}[b]{0.5\textwidth}
\resizebox{80mm}{!}{\includegraphics{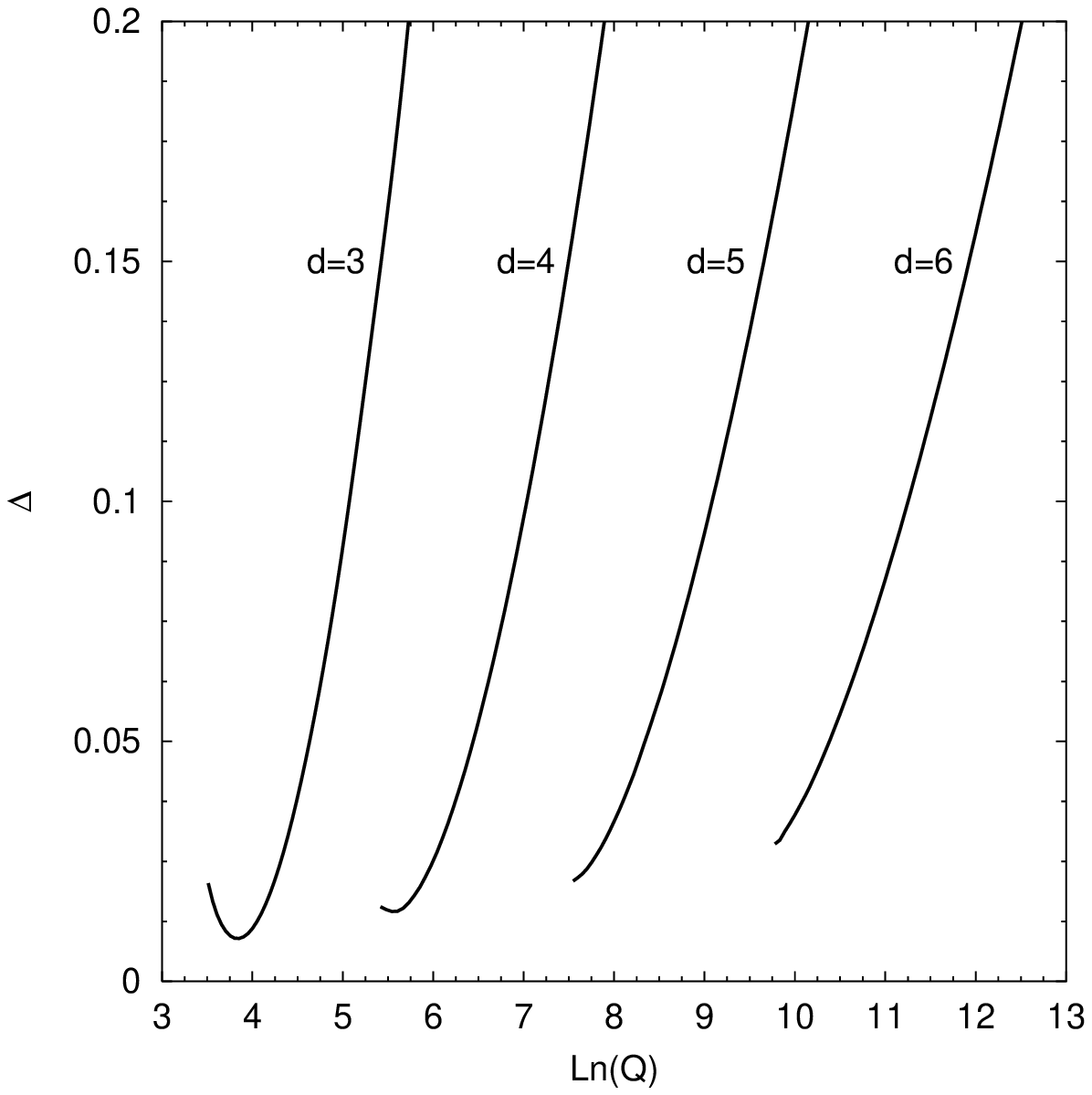}}
    \end{minipage}%
\caption {Contrasting numerical and analytical estimates for the
$Q$-ball energy for $d=3$, $4$, $5$, and $6$ (from left to right).
$b'=2.0 (2.3)$ on the left(right). It is clear that the gaussian
ansatz provides a good approximation for $Q$-balls with relatively
low charge.} \label{energyplotg}
\end{figure}

In fig. \ref{energyplotg} we contrast the analytical and numerical
values for the $Q$-ball energy as a function of $Q$ for $b'=2$
(left) and $b'=2.3$ (right) for $d=3,~4,~5$, and $6$. As for the
large-charge regime, there is a wide window of parameter space
well-described by the gaussian approximation. As the charge increases
the agreement gets worse. This is easily understood by examining the
``friction'' term in the equation for $\Phi(r)$, eq.
\ref{eqofmotion1}. As $Q$ increases the particle spends more
``time'' not moving ($\Phi'\simeq 0$, the motion is initially
friction-dominated) until $r$ grows large enough and the term
becomes subdominant. Since a gaussian profile does not approximate
well a flat core, the approximation gets worse for large $Q$
configurations. As before, we use $\Delta\equiv \left(E/Q|_{\rm num}
- E/Q|_{\rm anal}\right)/E/Q|_{\rm num}$.

\begin{figure}
\centering \resizebox{80mm}{!}{\includegraphics{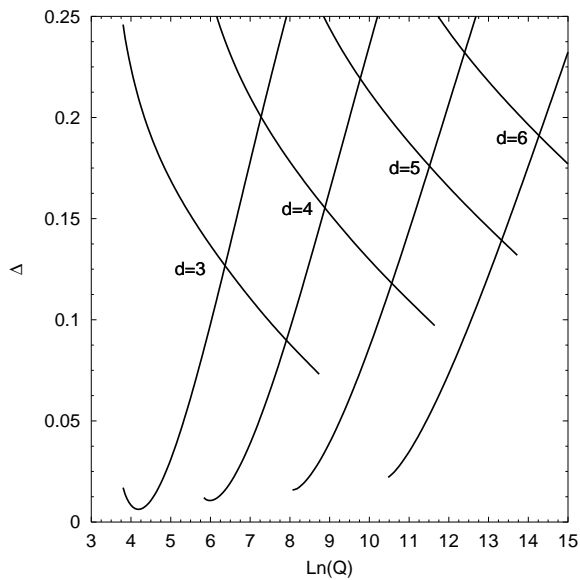}} \caption
{Contrasting the absolute errors of the variational estimates for
the energy of $Q$-balls for several values of spatial dimensionality
$d$. All plots are of $b'=2.0$. The positive slope line is for the
gaussian approximation while the negative slope line is for the
exponential-decay approximation of the previous section.}
\label{compgaussexp}
\end{figure}

We can now put the results of our analytical approximations together,
comparing their respective range of validity. In fig. \ref{compgaussexp}
we plot the percentual errors ($\Delta$)
for both approximations and $b'=2$. [As $b'\rightarrow 2.3$ one should use the gaussian
approximation as long as $Q$ is small enough, as can be seen in fig. \ref{energyplotg}.]
Note how the lines cross at a
given charge $Q_c$ for each value of $d$. Clearly, one is to use the
approximation with the smallest error: configurations
with $Q\leq Q_c$ are best approximated by the gaussian ansatz: they can be
called ``small'' $Q$-balls. Configurations
with $Q > Q_c$ are better described by the exponential ansatz of the previous section; they
can be called ``large'' $Q$-balls.
Choosing one or the other approximation, we are able to keep the errors quite small for different
values of $d$. For example, for $d =4$ and $b'\gtrsim 2$, the error can be kept below $15\%$.
This illustrates the validity of our analytical approximations and their
power to express the properties of $Q$-balls in an arbitrary number of
dimensions.

\begin{figure}
\centering \resizebox{80mm}{!}{\includegraphics{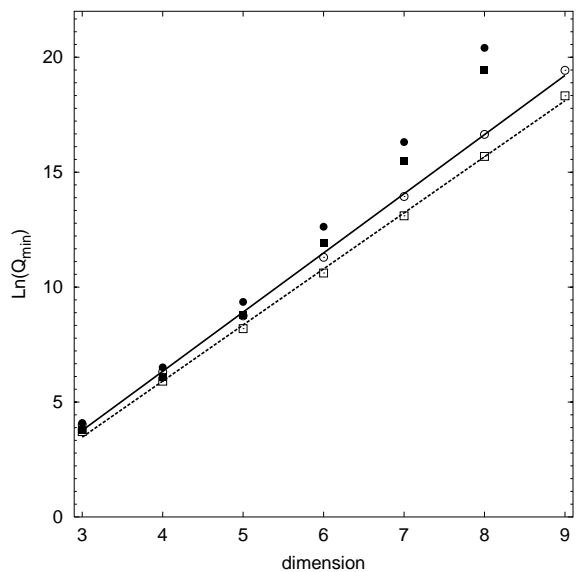}} \caption
{Minimum stable $Q$-ball charge vs. spatial dimensionality. The
circles (squares) represent $b'=2.0,(2.3)$ The filled dots are the
results from the gaussian approximation, eq. \ref{qmin}. The dashed
(solid) lines are exponential fits to the numerical values.}
\label{qminplot}
\end{figure}

As a final result,
we can use the gaussian ansatz to obtain an approximate expression for the lowest value for
the $Q$-ball charge, $Q_{\rm min}$. Imposing $E/Q=1$ on eq. \ref{energyg2}, one obtains a
polynomial identity, which is solved for

\begin{equation}
Q_{\rm min}^{2/d} = \frac{\xi}{2c}\left [1- \sqrt{1-\frac{4(d-2)^2c}{d}}\right ]~,
\label{qmin}
\end{equation}
where $c\equiv X_c/\sqrt{\gamma} - 2$. Note that for $d=2$ there is no restriction on the
minimum charge. That is not the case in higher dimensions.
We can use $X_c^2=3b'/4$ to obtain an expression of $Q_{\rm min}$ in terms
$b'$ and $d$. In fig. \ref{qminplot} we show $Q_{\rm min}$ as a function of $d$ for
$b' = 2.0$ (circles) and $b'=2.3$ (squares).
Filled symbols were computed using the gaussian ansatz above, while open symbols were obtained
numerically (cf. fig. \ref{numerQ}). The continuous lines are exponential fits
$Q_{\rm min}\sim \exp[d]$. $Q$-balls, even those with minimal charge, grow quite large in
higher dimensions, not a surprising result given that the charge is given by a volume
integral. As $d$ increases, the gaussian prediction for the minimum charge becomes worse
for reasons explained above.

\section{$d$-Dimensional $Q$-Clouds}

\begin{figure}
\centering \resizebox{80mm}{!}{\includegraphics{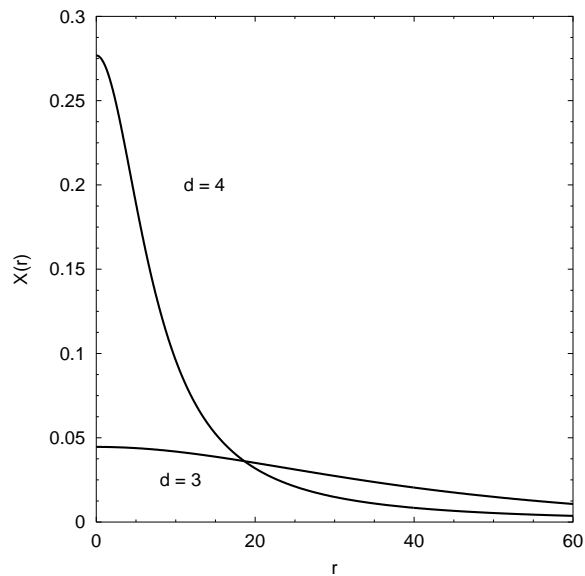}}
\caption {Q-cloud profiles for $d=4,3$ (upper,lower) for identical
parameters $b'=2.0,\omega=0.99999$.} \label{clouds}
\end{figure}

Q-clouds are classically unstable solutions ($E/Q>1$) that exist
as $\omega \rightarrow 1$ found for $d=3$ in ref. \cite{Q-clouds}. 
They belong to the upper branch of fig. \ref{numerQ}, starting at $Q_{\rm min}$,
characterized by low core field
values and large radii. Two examples can be seen in fig. \ref{clouds}.

It is quite simple to understand the qualitative nature of these solutions.
Recall, from eq. \ref{eqofmotion1}, that the solution traces the motion
of a ``particle'' in the potential $-U(\Phi)$, with a quadratic term,
$\frac{1}{2}(1-\omega^2)\Phi^2$, and a friction term $\frac{d-1}{r}\Phi'$. The
boundary condition at spatial infinity is, as for $Q$-balls, 
$\Phi\rightarrow 0$. As $\omega\rightarrow 1$, 
the hill at $U(0)$ gets shallower. Thus, it becomes easier
for the particle to reach $\Phi=0$. As a consequence, 
the value of the field at $r=0$,
$\Phi_c$, must decrease to avoid overshooting. The correct solution is obtained
when $\Phi_c$ is just large enough for the ``particle'' to overcome 
the friction and come to rest at $-U(\Phi=0)=0$ as $r\rightarrow \infty$. 

\begin{figure}
\centering \resizebox{80mm}{!}{\includegraphics{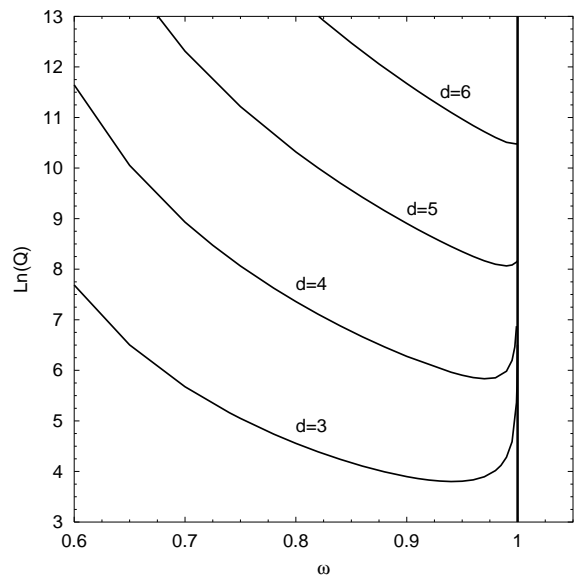}}
\caption {Q vs. $\omega$ for spatially-bound configurations in
different spatial dimensions and $b'=2$. Those with $\omega \leq
\omega_{min}$ are $Q$-balls, while those with $\omega >
\omega_{min}$ are $Q$-clouds. Note that $\omega_{min} \rightarrow 1$
for $d\geq 6$.} \label{omegavsQ}
\end{figure}

The transition from a $Q$-ball to a $Q$-cloud is controlled by the value of
$\omega$. Alford has shown that there is a critical value of $\omega$,
$\omega_{\rm min}$, above which the unstable $Q$-ball becomes a 
$Q$-cloud \cite{Q-clouds}. We have found that, indeed, 
the same behavior persists for $d=4$
and, marginally, for $d=5$. for $d=6$ we found that $\omega_{\rm min}\rightarrow
1$, as shown in fig. \ref{omegavsQ}. Clearly, as $d$ increases, and thus the
``friction,'' $\Phi_c$ must also increase. We illustrate this in fig. \ref{clouds}
where, for the same parameters ($b'$ and $\omega$), $\Phi_c$ grows 
substantially with $d$. Numerically, we were unable to find 
$Q$-clouds for $d\geq 6$. Apparently, in higher dimensions
only $Q$-balls, with larger values of
$\Phi_c$ to overcome the hill at $-U(0)$, are possible. Although we have not
proven this, we have determined that $Q$-clouds are only possible if
$\Phi_c < \Phi_{\rm inf}$, where $\Phi_{\rm inf}$ is the inflection
point nearest to the origin.

\section{Concluding Remarks}

We have presented a detailed study of $Q$-balls in an arbitrary number of
spatial  dimensions for potentials of the form
$V(\phi\phi^{\dagger})= \sum_{n=1}^{3} a_n(\phi\phi^{\dagger})^n$,
where $a_n$ is a constant.
Numerical solutions were obtained and contrasted with
simple analytical ansatze, separated into ``large'' and ``small'' $Q$-balls. 
It was shown that it is possible, for most of the parameter space, to use the
analytical expressions with comfortable accuracy to compute the $Q$-balls'
energy and radii as a function of only the spatial dimensionality and a single
parameter defining the potential energy density of the model.  We have also
obtained approximate analytical expressions for the minimum charge for
$Q$-balls to be stable, $Q_{\rm min}$, showing in addition how it depends on
spatial dimensionality. An empirical relation was found for the dependence of
the minimum charge on spatial dimensionality, $Q_{\rm min}\sim \exp[d]$.
Finally, we also explored the properties of $Q$-clouds in $d$ dimensions
for the same type of potentials. We
provided qualitative arguments as to why we were unable to find such solutions
for $d\geq 6$.

$Q$-balls have been studied with great interest since Coleman introduced them
in 1985. We hope that the present work will be useful to establish their possible
role in models that make use of extra spatial dimensions. By exploring the properties
of spatially-bound nonlinear objects in an arbitrary number of dimensions we gain
insight not only on the nature of such objects from a general viewpoint, but also
on the role of dimensionality on the properties of nonperturbative configurations.
Such objects, if existent, will no doubt have an impact 
on the physics of the vacuum and of the early universe, and may well influence the
dynamics of spatial compactification.

MG was supported in part by a National Science foundation Grant
no. PHYS-0099543.

\end{document}